\documentstyle[aps,prl,epsfig]{revtex}

\title{Resonances of the Frobenius-Perron Operator for a Hamiltonian
 Map with a Mixed Phase Space}

\author{Joachim Weber$^{1,2}$, Fritz Haake$^1$, Petr A. Braun$^{1,3}$, 
Christopher Manderfeld$^1$, and Petr \v{S}eba$^4$}
\address{$1$ Fachbereich Physik, Universit\"at-GH Essen, 45117 Essen, 
Germany}
\address{$2$ Department of Physics of Complex Systems, Weizmann 
Institute of Science, Rehovot 76100, Israel}
\address{$3$ Institute of Physics, Saint-Petersburg University, 
Saint-Petersburg 198504, Russia}
\address{$4$ Institute of Physics, Czech Academy of Sciences, Prague, 
Czech Republic}

\begin{document}

\draft
\date{Date: \today}
\maketitle
\begin{abstract}
Resonances of the (Frobenius-Perron) evolution operator ${\cal P}$ for
phase-space densities have recently attracted considerable attention, in
the context of interrelations between classical and quantum dynamics. We
determine these resonances as well as eigenvalues of ${\cal P}$ for
Hamiltonian systems with a mixed phase space, by truncating ${\cal
P}$ to finite size in a Hilbert space of phase-space functions and then
diagonalizing. The corresponding eigenfunctions are localized on
unstable manifolds of hyperbolic periodic orbits for resonances and on
islands of regular motion for eigenvalues. Using information drawn from
the eigenfunctions we reproduce the resonances found by diagonalization
through a variant of the cycle expansion of periodic-orbit theory and as
rates of correlation decay.
\end{abstract}

\section{Introduction}

Chaotic behavior in classical Hamiltonian systems manifests itself in an
infinite hierarchy of phase-space structures and instability of
trajectories with respect to small changes in their initial condition.
If resolution in phase space is restricted, e.g. due to finite precision
of measurements, phase-space trajectories can be followed only
approximately. Every finite-resolution initial condition actually stands
for an entire ensemble of trajectories with possibly very different
long-time behaviors. That fact suggests a probabilistic approach to the
dynamics, in which a phase-space density and its propagator ${\cal P}$,
the so called Frobenius-Perron operator, are studied.

Because of Liouville's theorem ${\cal P}$ can be represented by an
infinite unitary matrix in a Hilbert space of phase-space functions. The
spectrum of ${\cal P}$ thus lies on the unit circle in the complex
plane. Discrete eigenvalues represent regular dynamics, while chaos
entails a continuous spectrum, and along with the latter resonances of
${\cal P}$ may occur that characterize effectively irreversible
behavior\cite{Ruelle,Pollicott}. Apart from their role as decay rates, 
Frobenius-Perron
resonances have been found to link classical with quantum dynamics
because they could as well be identified from quantum systems
\cite{Us,Pance}. Furthermore, they have been predicted to carry
information on the system-specific corrections to universality of
quantum fluctuations\cite{Altshuler,Zirnbauer}.

To identify resonances, mathematically defined as singularities of the
resolvent of ${\cal P}$ in some higher Riemannian sheet, it is necessary to
analytically continue the resolvent across the continuous spectrum of
${\cal P}$ from the outside to the inside of the unit circle
\cite{Reed,Fishman,Khodas}. This becomes increasingly difficult as the 
dynamics gains complexity. 
Especially for the large class of systems with a mixed phase
space an effective approximation scheme is needed that is free of
restrictions of previous investigations \cite{Hasegawa,Baladi}, such as
hyperbolicity, one-dimensional (quasi-) phase space, or isolation of the
phase-space regions causing intermittency.

In the following we present such a scheme and illustrate it for a
prototypical dynamical system with a mixed phase space, a periodically
kicked top. We represent ${\cal P}$ in a Hilbert space of phase-space
functions as an infinite unitary matrix and subsequently truncate it to
a finite $N$-dimensional matrix ${\cal P}^{(N)}$, with resolution in
phase space as the truncation criterion. We thus generate a nonunitary
approximation to the unitary ${\cal P}$ that becomes exact as
$N\to\infty$. Looking at the spectra of ${\cal P}^{(N)}$ with increasing
$N$, we find some eigenvalues persisting in their positions, either
almost on the unit circle or well inside, and we focus on these
``frozen'' eigenvalues. While frozen (near)unimodular eigenvalues of
${\cal P}^{(N)}$ turn into unimodular eigenvalues of ${\cal P}$ as
$N\to\infty$, the nonunimodular frozen eigenvalues are not part of the
spectrum of ${\cal P}$ in that limit. We will argue that at finite but
large $N$ they rather indicate the positions of resonances of ${\cal
P}$. Accordingly, the eigenfunctions of ${\cal P}^{(N)}$ are of very
different nature for these two cases. We find eigenfunctions to
unimodular eigenvalues localized and approximately constant on islands
of regular motion in phase space that are bounded by invariant tori. In
contrast, eigenfunctions to nonunimodular eigenvalues of ${\cal
P}^{(N)}$ are localized near unstable manifolds of hyperbolic periodic
orbits. Just as nonunimodular eigenvalues are not in the spectrum of the
Hilbert-space operator ${\cal P}$, the eigenfunctions are no
Hilbert-space functions in the limit of infinite resolution. A
comparison of the eigenfunctions of ${\cal P}^{(N)}$ for different
values of $N$ and furthermore with the eigenfunctions of the truncated
inverse operator ${\cal P}^{-1 (N)}$ illustrates why they lie outside
the Hilbert space. The strong localization of these
(resonance-)eigenfunctions at finite $N$ allows us to identify the
relatively few groups of periodic orbits associated with a particular
resonance up to a certain length. From only these orbits we recover
resonances via a variant of the so-called cycle expansion of
periodic-orbit theory that works even in the case of a mixed phase
space. To verify that frozen eigenvalues have indeed the physical
interpretation of resonances, we conclude with a numerical experiment in
which the decay of a two-point correlation function is observed. In
order to find a certain resonance responsible for correlation  decay,
the initial phase-space density is chosen localized in the phase-space
region where the resonance-eigenfunction has large amplitudes.

A short account of the results to be discussed here has been published
recently \cite{Weber}.

\section{Liouville dynamics on the sphere}

As a prototypical Hamiltonian system with a phase space displaying a mix 
of chaotic and regular behavior we consider a periodically kicked angular 
momentum vector 
\begin{equation}
(J_x,J_y,J_z) = (j \sin \theta \cos
\varphi, j \sin \theta \sin \varphi, j \cos \theta)
\label{eq1}
\end{equation}
of conserved length $j$, also known as the kicked top \cite{Haake}. Such a
system has one degree of freedom, and its phase space is a sphere. A
phase-space point $X\equiv (p,q)$ may be characterized by an
``azimutal'' angle $\varphi$ as the coordinate $q$ and the
cosine of a ``polar'' angle $\theta$ as the conjugate momentum $p$. The
dynamics is specified as a stroboscopic area preserving map, $X'=M(X)$.
We chose $M$ to consist of rotations $R_z(\beta_z), R_y(\beta_y)$
about the $y$- and $z$-axes and a ``torsion'', i.e. a nonlinear rotation
$T_z(\tau) = R_z(\tau \cos \theta)$ about the $z$-axis which changes
$\varphi$ by $\tau \cos \theta$,
\begin{equation}
M= T_z(\tau) R_z(\beta_z) R_y(\beta_y) \, .
\label{eq2}
\end{equation}
In the sequel we keep $\beta_z = \beta_y = 1$ fixed and vary the torsion
constant $\tau$ to generate a phase space which is integrable for $\tau
= 0$ and increasingly dominated by chaos as $\tau$ grows. We focus on
$\tau = 2.1$ as a moderately chaotic case and $\tau=10.2$ as strongly
chaotic (see figures \ref{figure1}a,b), where elliptic islands have
become so small that they are difficult to detect.

In the Liouville picture the time evolution of a phase-space density $\rho$
is governed by Liouville's equation,
\begin{equation}
\partial_t \rho = {\cal L} \rho = \left\{H ,\rho\right\},   
\label{eq3}
\end{equation}
where the Liouville operator ${\cal L}$, the Poisson bracket with the
Hamiltonian $H$, appears as generator. For our rotations and torsion we
separately take $H = \beta_y J_y, H = \beta_z J_z$, and $H = (\tau/2) 
J_z^2$. Denoting the respective Liouvillians by ${\cal L}_{R_y}$, 
${\cal L}_{R_z}$, and ${\cal L}_{T_z}$ we imagine Liouville's equation for
each of them integrated over a unit time span. The product of the
resulting three propagators yields our Frobenius-Perron operator ${\cal
P}=\exp({\cal L}_{T_z})\exp({\cal L}_{R_z})\exp({\cal L}_{R_y})$. The
action of ${\cal P}$ on the phase-space density $\rho(q,p)$ can be
represented in terms of the "Hamilton-picture" map $M$ as
\begin{eqnarray}
\rho_{n+1}(X) & = & {\cal P} \;\rho_{n}(X)=\,
{\rm e}^{\,{\cal L}_{T_z}}{\rm e}^{\,{\cal L}_{R_z}}{\rm e}^{\,{\cal
L}_{R_y}}\;\rho_{n}(X) \nonumber\\
& = & \int {\rm d}X'\, \delta(X-M(X'))\; \rho_n(X')\; .
\label{eq3b}
\end{eqnarray}
Since the map $M$ is invertible and area preserving, $\det
\partial M(X)/\partial X=1$, the integral simplifies to
\begin{equation}
{\cal P} \rho(X) = \rho\left[M^{-1}(X)\right] \, .
\label{eq4}
\end{equation}

While a phase-space density is usually considered as $L^1$-integrable it
can be assumed to belong to a Hilbert space of $L^2$-functions as well.
The additional structure of the Hilbert space can be exploited to
represent ${\cal P}$ by an infinite unitary matrix, the unitarity
$({\cal P}\rho_1,{\cal P}\rho_2)=(\rho_1,\rho_2)$ following from
(\ref{eq4}) and area preservation in the map $M$. The basis functions
may be chosen as ordered with respect to phase-space resolution. This
allows us to eventually truncate ${\cal P}$ to finite size and still
retain the propagator intact for functions on large phase-space scales.

On the unit sphere the spherical harmonics $Y_{lm}(\varphi,\cos\theta)$ with
$l = 0, 1, 2, \dots$ and integer $|m|\le l$ form a suitable basis.
Phase-space resolution is characterized by the index $l$: using all $Y_{lm}$
with $0 \le l \le l_{{\rm max}}$ phase-space structures of area  $\propto
1/l_{{\rm max}}^2$ can be resolved, and the number of these basis
functions is $N=(l_{{\rm max}}+1)^2$. 

To check on the resolution achievable by spherical harmonics consider a 
Gaussian $G(\varphi,\theta)=\exp{( -{\rm i}(\varphi-\varphi_0)^2/
\Delta\varphi^2-{\rm  i}(\theta-\theta_0)^2/\Delta\theta^2)}$ with the 
angular widths $\Delta\varphi,\Delta\theta$ and the effective area 
$\sim\Delta\varphi\Delta\theta$. To represent it by a linear combination 
of spherical harmonics we need only the $Y_{lm}$ with $|m|\leq m_0\sim 
2\pi/\Delta\varphi$ and $l\leq 2\pi/\Delta\varphi+\pi/\Delta\theta$; 
all other spherical harmonics have negligible scalar product with 
$G(\varphi,\theta)$. Indeed, the effective ``wavelengths'' 
of $Y_{lm}$ in $\varphi$ and $\theta$ are, respectively, $2\pi/|m|$ and 
$\pi/(l-|m|)$, the latter given by the number $(l-|m|)$ of zeros of the 
Legendre function $P_{lm}$ in the interval $0<\theta<\pi$; once these 
wavelengths become smaller then the respective widths $\Delta\varphi,
\Delta\theta$ of the Gaussian the overlap of $G$ and $Y_{lm}$ is 
negligible. Inverting that statement we may say that a basis set of 
spherical harmonics cut off at a certain $l_{\rm max}$ is suitable for 
representing objects on the unit sphere with the angular sizes 
$\sim l_{\rm max}$ and area $\sim l_{\rm max}^2$.

To obtain the matrix elements 
\begin{equation}
{\cal P}_{l m, l' m'} = \int_{-1}^1 {\rm d}\cos\theta\; \int_0^{2 \pi} 
{\rm d}\varphi \; Y_{lm}^\ast(\varphi,\cos\theta) Y_{l'm'}
\left[M^{-1}(\varphi,\cos\theta)\right]
\label{eq5}
\end{equation}
for the mixed dynamics it is convenient to first determine the
Frobenius-Perron matrices for the two rotations $\exp({\cal L}_{R_z}),
\exp({\cal L}_{R_y})$ and the torsion $\exp({\cal L}_{T_z})$, all of
which are by themselves integrable and free of chaos. Deferring the
derivation to Appendix A we here just note the three matrices. The
Frobenius-Perron matrix of the $z$- rotation is diagonal,
\begin{equation}
\Big(\exp({\cal L}_{R_z})\Big)_{ l m, l' m'} =
\delta_{l l'} \delta_{m m'} \exp(-{\rm i} m \beta_z) \, .
\label{eq9}
\end{equation}
For the $y$-rotation, $(\exp({\cal L}_{R_y}))_{ l m, l' m'}$ is 
blockdiagonal in $l$ with the blocks well known from quantum mechanics 
as Wigner's ${\rm d}$-matrices,
\begin{equation}
\Big(\exp({\cal L}_{R_y})\Big)_{ l m, l' m'} = \delta_{l l'} 
{\rm d}^l_{m m'}(\beta_y)\,.
\label{eq10}
\end{equation}
For the $z$-torsion the Frobenius-Perron matrix elements are given by
finite sums over products of spherical Bessel functions $j_l(x)$ and
Clebsch-Gordan coefficients,
\begin{eqnarray}
\Big(\exp({\cal L}_{T_z})\Big)_{l m, l' m'}  & = &
\delta_{m, m'} (-1)^{m} \sqrt{(2l+1)(2l'+1) } \nonumber\\ \times
& & \sum_{l''=|l-l'|}^{l+l'}  (-{\rm i})^{l''} \; j_{l''}(m\tau) \;
C_{0\, 0\, 0}^{l\,\, l'\,\, l''} \;
C_{-m\, m\, 0}^{l\,\, l'\,\, l''}\,.
\label{eq11}
\end{eqnarray}
Finally, for the mixed dynamics ${\cal P}$ is the product of the three
above matrices. Due to the Kronecker deltas involved, each element of
${\cal P}$ is just the product of three matrix elements, one each for the
rotations and the torsion. Obviously, ${\cal P}$ is a full matrix
coupling all phase-space scales, as it is to be expected for chaotic
dynamics. We furthermore remark that with respect to the  real basis
\begin{eqnarray}
y_{l0} & = & Y_{l0} \nonumber\\
y_{lm}^+ & = & \frac{1}{\sqrt{2}} (Y_{lm} + Y_{lm}^*) \, , \\
y_{lm}^- & = & \frac{1}{\sqrt{2} {\rm i}} (Y_{lm} - Y_{lm}^*) \nonumber
\end{eqnarray}
the matrix ${\cal P}$ is also real  , since real phase-space densities 
cannot become complex under time evolution. As a consequence, all 
eigenvalues of ${\cal P}^{(N)}$ are either real or come in complex 
conjugate pairs.

\section{Resonances and eigenvalues at finite phase-space resolution}

To study resonances and (discrete) eigenvalues of ${\cal P}$ we investigate 
the finite-size approximants ${\cal P}^{(N)}$ for different values of $N$, 
i.e. different resolutions in phase space. In the low-resolution subspace 
spanned by the first $N$ basis functions ${\cal P}^{(N)}$ replaces 
${\cal P}$ identically, but it completely rejects fine phase-space 
structures that cannot be expanded in terms of these $N$ functions.

The truncation of ${\cal P}$ to finite size breaks unitarity. Therefore,
${\cal P}^{(N)}$ may   not only have unimodular
eigenvalues but also eigenvalues located inside the unit circle.
Furthermore, the spectrum of ${\cal P}^{(N)}$ is purely discrete.
The $N$-dependence of the eigenvalues of ${\cal P}^{(N)}$ yields
information on the spectral properties of ${\cal P}$. Upon diagonalizing
${\cal P}^{(N)}$ and increasing $N$ we find the ``newly born''
eigenvalues close to the origin while the ``older'' ones move about in
the complex plane. ``Very old'' ones eventually settle for good. If the
classical dynamics is integrable ($\tau = 0$ or $\beta_y=0$), the
asymptotic large-$N$ loci are back to the unit circle, where the full
${\cal P}$ has its discrete eigenvalues. The situation is different for 
the mixed phase space: while some eigenvalues of ${\cal P}^{(N)}$ 
``freeze'' with unit moduli, others come to rest inside the unit circle 
as $N \to \infty$. These eigenvalues reflect resonances of ${\cal P}$ 
in a higher Riemannian sheet of the complex plane. We see this phenomenon 
as analogous to the spectral concentration in perturbation theory
\cite{Titchmarsh,Reed,Kato}: a perturbation series for an operator with
continuous spectrum does not converge but produces, with increasing
order, a sequence of points concentrated in the neighborhood of the
respective resonance.

An additional intuitive argument for the persistence of non-unimodular
eigenvalues as $N\to\infty$ for nonintegrable dynamics is the
following: in contrast to regular motion, chaos comes with a hierarchy
of phase-space structures which extends without end to ever finer
scales. Probability that is propagated from large to fine scales mostly
does not return. This effectively dissipative character of the
conservative dynamics is accounted for as true dissipation by the
truncated propagator and its nonunimodular eigenvalues, but the
difference between effective and true dissipation is irrelevant for low
resolution. Moreover, a frozen nonunimodular eigenvalue is to be
understood as a scale-independent dissipation rate, in tune with the
expected selfsimilarity in phase space.

\begin{figure}
\epsfxsize=0.4\textwidth
\epsffile{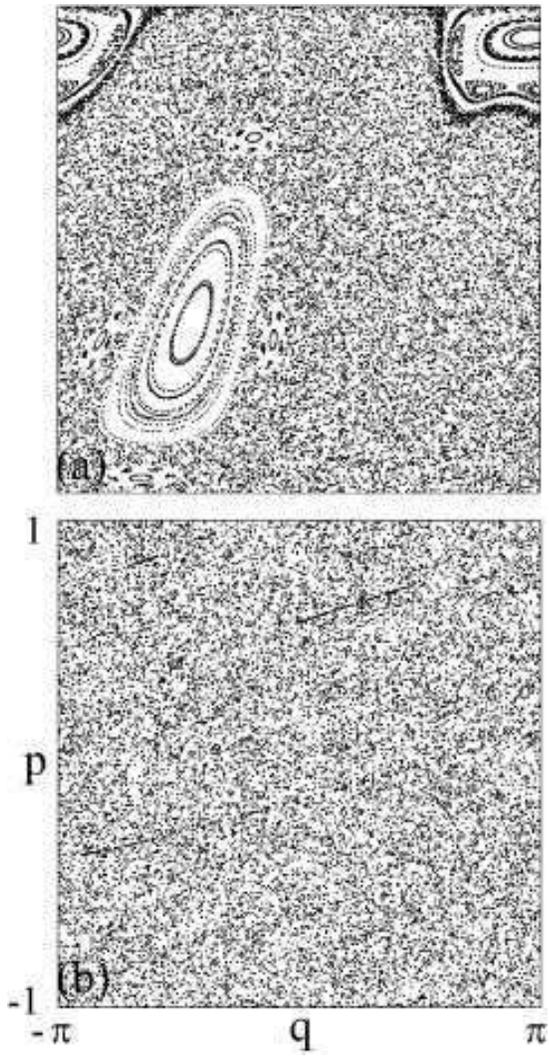}
\caption{Phase-space portraits of the weakly chaotic kicked top with 
torsion parameter $\tau = 2.1$ (a), and of the strongly chaotic top 
with $\tau = 10.2$ (b).}
\label{figure1}
\end{figure}

\begin{figure}
\epsfxsize=0.4\textwidth
\epsffile{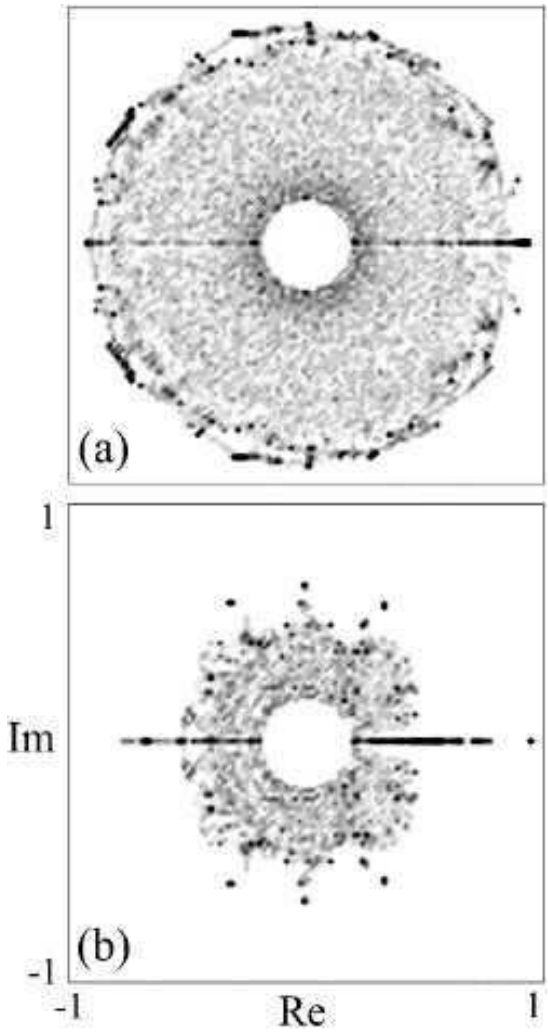}
\caption{Histograms showing the distributions of eigenvalues with 
moduli larger than $0.2$ of all matrices with $l_{{\rm max}} = 40,41,
\ldots,70$ in the complex plane for $\tau = 2.1$ (a) and $\tau=10.2$ (b). 
Large amplitudes (black) indicate positions of frozen eigenvalues.}
\label{figure2}
\end{figure}

The freezing of non-unimodular eigenvalues is illustrated for two kicked
tops, one with weakly chaotic $(\tau = 2.1)$ and one with strongly
chaotic character $(\tau = 10.2)$. Figures \ref{figure1}a and
\ref{figure1}b show the respective phase-space portraits. While for
$\tau = 2.1$ large islands of regular motion still exist, for $\tau =
10.2$ regular islands have become so small that they are difficult to
detect and impossible to resolve with $l_{{\rm max}} = 70$. Figures
\ref{figure2}a and \ref{figure2}b show grey-shade coded histograms in
the complex plane for the number of eigenvalues with moduli larger than
$0.2$ of all matrices with $l_{{\rm max}} = 40,41,42, \dots 70$. 
Eigenvalues with moduli smaller than $0.2$ are rejected because they 
have not settled yet and their density near the origin is so high that 
they would spoil the histogram. Large amplitudes (black) in the 
histograms indicate the positions of frozen eigenvalues. While for 
$\tau=2.1$ a large number of frozen eigenvalues is found close to the 
unit circle, in the case $\tau = 10.2$, where no regular islands can 
be resolved, the only unimodular eigenvalue is the one at unity which 
pertains to the stationary uniform eigenfunction $Y_{00}$; all other 
frozen eigenvalues lie well inside the unit circle and indicate 
resonance positions. Table \ref{tableA} lists some of these frozen 
nonunimodular eigenvalues at resolutions $l_{{\rm max}} = 30, 40, 50$ 
and $60$.

At first sight the two cases presented above may appear to correspond to
quite different types of dynamics, but actually this is only a matter of
resolution. As resolution is further increased in the strongly chaotic
case $(\tau = 10.2)$, elliptic islands will be resolved at some stage.
Correspondingly, eigenvalues of ${\cal P}^{(N)}$ will emerge which
freeze on the unit circle. Also new nonunimodular frozen eigenvalues may
appear, possibly even ones larger in modulus than the frozen eigenvalues 
found at lower resolution.

\section{Eigenfunctions and resonance-eigenfunctions at finite phase-space 
resolution}

\begin{figure}
\epsfxsize=0.32\textwidth
\epsffile{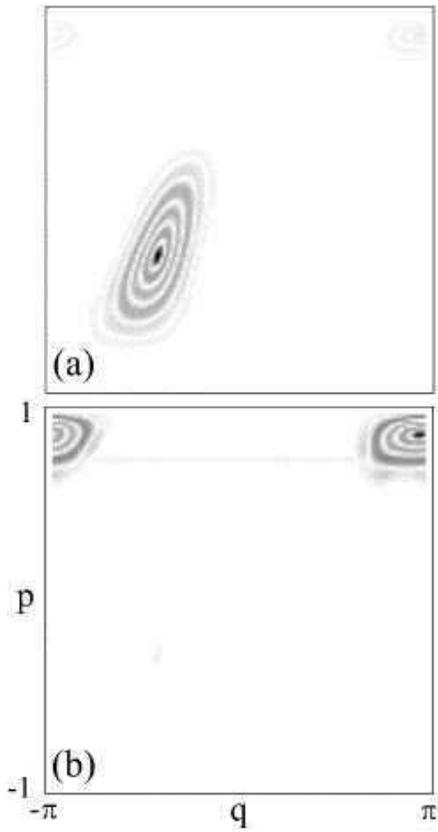}
\caption{Eigenvalues $0.999976$ (a) and $0.999974$ (b) almost at unity 
of ${\cal P}^{(N)}$ with $l_{{\rm max}} = 60$ in the weakly chaotic case 
$\tau=2.1$ have eigenfunctions that are localized on elliptic islands.
Dark-shaded regions in phase space indicate large amplitudes of the 
eigenfunctions.}
\label{figure3}
\end{figure}

An investigation of the eigenfunctions of ${\cal P}^{(N)}$ reveals more
about the nature of the frozen eigenvalues. Eigenvalues freezing with
unit moduli have eigenfunctions located on elliptic islands of regular
motion surrounding elliptic periodic orbits in phase space. Such islands
are bounded by invariant tori which form impenetrable barriers in phase
space. Thus if $\chi_{p, n}$ is a characteristic function constant on an
island around the $n$th point of an $p$-periodic orbit and zero outside
with $\chi_{p, n+1} = {\cal P} \chi_{p, n}$ and $\chi_{p, 1} = {\cal P}
\chi_{p, p}$, then
\begin{equation}
f_p = \sum_{n=1}^p \chi_{p, n}
\end{equation}
is an eigenfunction of ${\cal P}$ with eigenvalue unity.
Moreover, linear combinations of such functions are eigenfunctions as
well. Figures \ref{figure3}a,b show two eigenfunctions of ${\cal
P}^{(N)}$ with eigenvalues $0.999976$ and $0.999974$ almost at unity for 
the kicked top with $\tau = 2.1$ and $l_{{\rm max}} = 60$. Amplitudes of 
the functions are large in dark shaded areas of phase space. The elliptic 
islands supporting the eigenfunctions can easily be identified in the 
phase-space portrait in figure \ref{figure1}a.

For $p>1$ we furthermore expect and find the $p$th roots of unity as
eigenvalues as well with eigenfunctions that have constant moduli on
elliptic islands and are invariant under ${\cal P}^p$,
\begin{equation}
f_{p,m} = \sum_{n=1}^p {\rm e}^{{\rm i} 2 \pi m n/p} \chi_{p, n} \; , 
\quad \mbox{with} \; m = 1, \dots, p\; . 
\end{equation}
Superpositions of two such $f_{p,m}, f_{p',m'}$ are also 
eigenfunctions if $m/p = m'/p'$, i.e. if all periods $p'$ involved
are multiples of $p$; the possible phases are dictated by the
shortest period $p$.

\begin{figure}
\epsfxsize=0.32\textwidth
\epsffile{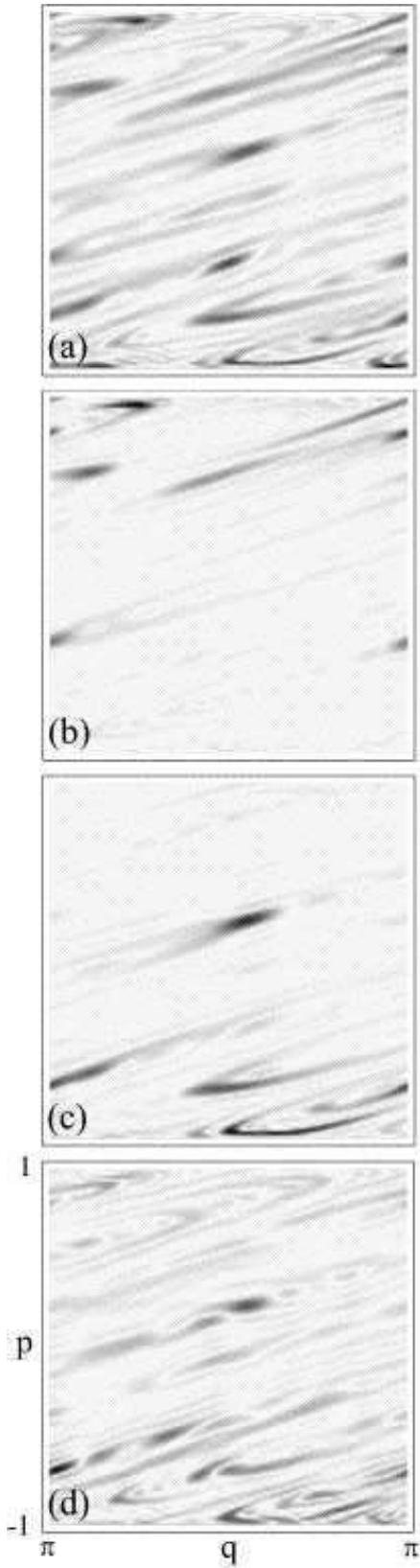}
\caption{For frozen eigenvalues $0.7696$ (a), $-0.3388\pm{\rm i}0.6243$ (b), 
$-0.0058\pm{\rm i}0.7080$ (c), $0.6480$ (d) of ${\cal P}^{(N)}$ 
($\tau=10.2$, $l_{{\rm max}} = 60$) inside the unit circle eigenfunctions 
have large amplitudes (dark-shaded areas) around unstable manifolds of 
hyperbolic periodic orbits in phase space.} 
\label{figure4}
\end{figure}

\begin{figure}
\epsfxsize=0.32\textwidth
\epsffile{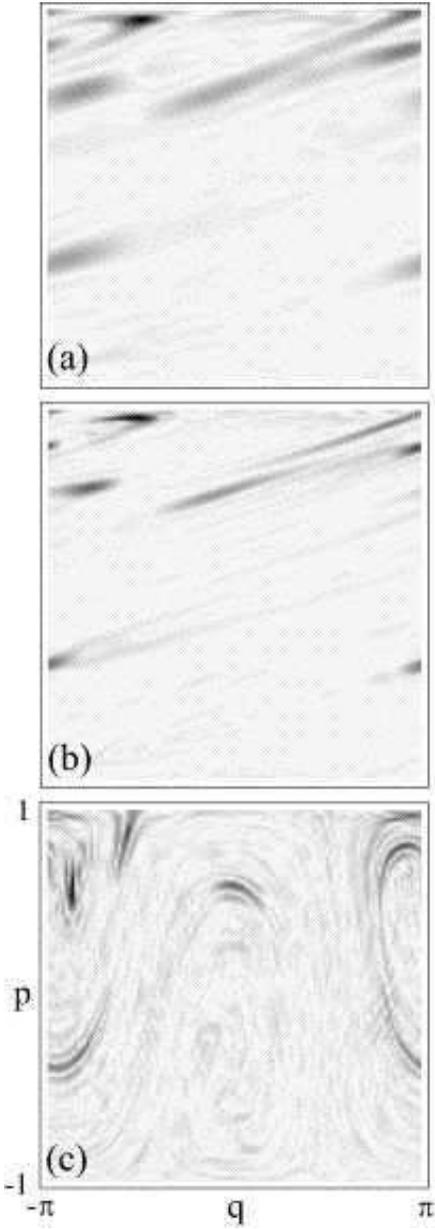}
\caption{As phase-space resolution increases from $l_{{\rm max}}=30$ (a) 
to $l_{{\rm max}}=60$ (b), the eigenfunction of ${\cal P}^{(N)}$ for the 
frozen nonunimodular eigenvalue  $-0.3388\pm {\rm i} 0.6243$ gains  new 
structures on finer scales $(\tau=10.2)$. The corresponding eigenfunction 
(c) of ${\cal P}^{-1 (N)}$  (resolution $l_{{\rm max}} = 60$) is localized 
at the same periodic orbit  as the eigenfunction (b) but with stable and 
unstable manifolds interchanged. Therefore, see also figure 
\protect\ref{figure6}a for the periodic orbits, figure 
\protect\ref{figure6}b for the unstable and figure \protect\ref{figure6}c 
for the stable manifolds; these manifolds are clearly dominating the 
present eigenfunction.}
\label{figure5}
\end{figure}

\begin{figure}
\epsfxsize=0.9\textwidth
\epsffile{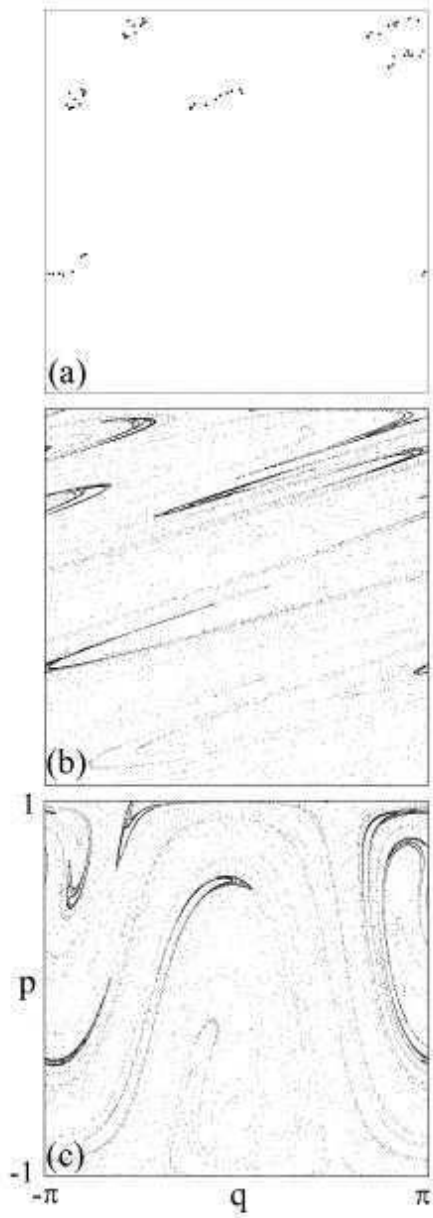}
\caption{(a): The $12$ orbits of primitive length $6$ that can be 
identified for the resonances $0.3550\pm {\rm i} 0.6199$ and 
$-0.3388\pm {\rm i}0.6243$ with phases near the sixth roots of unity 
from the eigenfunction shown in figures \protect\ref{figure4}b 
(also \protect\ref{figure5}b). They contribute to the expansion of 
${\rm d}_i(z)$ in lowest order. The unstable manifolds of these orbits 
are shown in (b), the stable manifolds in (c).}  
\label{figure6}
\end{figure}

Once freezing has been observed for eigenvalues with moduli smaller than
unity their eigenfunctions have approached their final shape on the
resolved phase-space scales. In contrast to the eigenfunctions with
unimodular eigenvalues these eigenfunctions are localized around
unstable manifolds of hyperbolic periodic orbits, ones with low periods
at first since these are easiest to resolve. 
[Typically, for every period several neighbouring orbits of approximately
equal instability can be identified in an eigenfunction, their number being
the larger the more extended the eigenfunction.] Figure \ref{figure4}
shows moduli of eigenfunctions for some of the nonunimodular eigenvalues
listed in table \ref{tableA} at resolution $l_{{\rm max}}= 60$.
Particularly in figures \ref{figure4}b,c the localization on orbits of
length six and four respectively is obvious. With growing $l_{{\rm
max}}$ more orbits of higher periods appear in the eigenfunction. Their
supports gain complexity, in correspondence with the infinitely
convoluted shape of the unstable manifolds. Just as for the eigenvalues
there is no strict convergence of the eigenfunctions. Since no finite
approximation ${\cal P}^{(N)}$ accounts for arbitrarily fine structures
one encounters the aforementioned loss of probability from resolved to
unresolved scales. Not even in the limit $N\to\infty$ can the unitarity
of ${\cal P}$ be restored: rather, in tune with a continuous spectrum of
$\cal P$, the eigenfunctions tend to singular objects outside the
Hilbert space. This becomes even more obvious in a comparison of
eigenfunctions of ${\cal P}^{(N)}$ with eigenfunctions of the truncated
inverse propagator ${\cal P}^{-1 (N)}$. Both matrices have the same
frozen eigenvalues, but only for the unimodular ones eigenfunctions of
${\cal P}^{(N)}$ and ${\cal P}^{-1 (N)}$ to the same eigenvalue
have (roughly) the same support, as it is expected for unitary dynamics: 
the elliptic islands that support the eigenfunctions are invariant under 
time reversal. For nonunimodular frozen eigenvalues eigenfunctions of 
${\cal P}^{(N)}$ and ${\cal P}^{-1 (N)}$ have distinct supports. 
While unstable periodic orbits are traversed in the opposite direction but 
remain the same under time reversal, stable and unstable manifolds are 
interchanged. Thus the eigenfunctions of ${\cal P}^{-1 (N)}$ are 
localized near the same periodic orbits but around the stable manifolds 
of the forward-dynamics. 
In particular the eigenfunctions do not converge to the same limit as
$N\to\infty$. Needless to say this again reflects the effective
irreversibility of the chaotic Hamiltonian dynamics. Figures
\ref{figure5} and \ref{figure6} illustrate the aforesaid. The eigenfunction
in figure \ref{figure5}a at resolution $l_{{\rm max}}=30$ corresponds 
to the same frozen eigenvalue as the eigenfunction in figure 
\ref{figure5}b at resolution $l_{{\rm max}}=60$. While coarse structures 
are the same in both eigenfunctions, more fine structures appear at 
the higher resolution. 
Figure \ref{figure5}c shows, again for the same eigenvalue,
the eigenfunction of ${\cal P}^{-1 (N)}$ at $l_{{\rm max}}=60$. In
contrast to figure \ref{figure5}b localization is now on the stable
manifolds of the same periodic orbits of the foreward dynamics.
These periodic orbits as well as their unstable and stable manifolds are
shown in figure \ref{figure6}.

Just like the unimodular eigenvalues, frozen eigenvalues inside the unit
circle come in $p$-families, i.e. with phases corresponding to the
$p$th roots of unity, determined by the length $p$ of the shortest 
periodic orbit present in an eigenfunction. Now the above argument must 
be modified in the following way: assume an eigenfunction $f$ is mostly 
concentrated around a shortest unstable orbit with period $p$ as well 
as a longer one with period $p'$. Denote by $\psi_{p,n}$ again a 
``characteristic function'', now constant near the $n$-th point of the 
period$-p$ orbit, $n=1\ldots p$, and zero elsewhere. The truncated 
Frobenius-Perron operator ${\cal P}^{(N)}$ maps $\psi_{p,n}$ into 
${\cal P}^{(N)}\psi_{p,n}=r_p\psi_{p,n+1}$ with the real positive factor 
$r_p$ smaller than unity accounting for losses, in particular to 
unresolved scales. 
As previously, independent linear combinations of the $\psi_{p,n}$ 
can be formed as
\begin{equation}
f_{p,m}=\sum_{n=1}^p{\rm e}^{{\rm i}2\pi m n/p}\psi_{p,n}\quad
\mbox{with}\; m =1\ldots p\, .
\label{eq21}
\end{equation}
Now consider a sum of two such functions, $g=f_{p,m}+f_{p',m'}$, and
apply ${\cal P}^{(N)}$. For $g$ to qualify as an approximate
eigenfunction we must obviously have $r_p\approx r_{p'}$, i.e. the two
periodic orbits must be similarly unstable, and $m/p=m'/p'$. But then
indeed ${\cal P}^{(N)}g\approx r_p{\rm e}^{{\rm i}2\pi m/p}g$ and
$[{\cal P}^{(N)}]^pg\approx r_p^pg$, and the phase is dictated by the
shortest orbit. Since orbits of low period are most likely to be
resolved first, usually the first frozen eigenvalues found from
diagonalizing ${\cal P}^{(N)}$ have phases corresponding to small
numbers $p$. Since nonunimodular eigenvalues with phases according to $p
= 1,2,4$ have already been presented in \cite{Weber}, in this paper we
intentionally chose $\tau =10.2$, a case for which eigenvalues with
other phases happen to be more easily detected. In figure \ref{figure2}b
and table \ref{tableA} complex eigenvalues with phases near the sixth
roots of unity are clearly visible.

\section{Periodic-orbit expansion of the spectral determinant}

Since the unstable periodic orbits that are linked to a nonunimodular
eigenvalue can be identified quite easily from the eigenfunction of
${\cal P}^{(N)}$, one is tempted to adopt a cycle expansion to calculate
decay rates from periodic orbits.

For purely hyperbolic systems, cycle expansions of the spectral
determinant, i.e. the characteristic polynomial of the Frobenius-Perron
operator
\begin{equation}
{\rm d}(z) = \det(1-z{\cal P})=\exp {\rm tr}\ln(1-z{\cal P})
\end{equation}
are known to allow for the calculation of
resonances with high accuracy \cite{Cvitanovic,Christiansen}.
The spectral determinant is expressed in terms of the traces of the
Frobenius-Perron operator ${\rm tr}\;{\cal P}^{n}$ as
\begin{equation}
{\rm d}(z) = \prod_{n=1}^{\infty} \exp\left(-\frac{z^n}{n} {\rm tr}\;
{\cal P}^n \right)
\end{equation}
and subsequently expanded as a finite polynomial up to some order
$n_{{\rm max}}$. Only the first $n_{{\rm max}}$ traces are required for
the calculation of this polynomial, and only periodic orbits with
primitive periods up to $n_{{\rm max}}$ contribute to it.

The traces  
\begin{equation}
{\rm tr}\; {\cal P}^n = \int {\rm d}X \; \delta(X-M^n(X))
\end{equation}
are calculated as sums over
hyperbolic periodic orbits of length $n$ as 
\begin{equation}
{\rm tr}\; {\cal P}^n=\sum_{X= M^n X} \frac{1}{|\det(1-\Xi_n)|}
\end{equation}
where the matrix $\Xi_n=\partial M^n(X)/\partial X$ is the
linearized version of the map $M^n$ evaluated at any of the points of a
period$-n$ orbit. If $n$ is the $r$-fold of a primitive period
$p$, the matrix $\Xi_n$ can as well be written as $\Xi_n = \Xi_p^r$, and
the spectral determinant as a product of contributions from all primitive
orbits $\{k\}$,
\begin{equation}
{\rm d}(z) = \prod_{\{k\}} \exp \left[ - \sum_{r=1}^\infty \frac{1}{r}
\frac{z^{p_k r}}{\det(1-\Xi_{p_k}^r)} \right] \,.
\end{equation}
The zeros $z_i$ of the polynomial which are insensitive against an 
increase of $n_{{\rm max}}$ are inverses of resonances.

For the ordinary cycle expansion of a spectral determinant just sketched
to converge, all periodic orbits must be hyperbolic and sufficiently
unstable \cite{Ruelle,Cvitanovic,Christiansen}. These conditions are
not met for a mixed phase space, not only because of the presence of
elliptic orbits but also because of very weakly unstable orbits with
high periods. We can circumvent such limitations by considering only one
ergodic region in phase space at a time, bar contributions from elliptic
orbits, and impose a stability bound by including only the relatively
few, about equally unstable hyperbolic orbits $\{k\}_i$ identified in an
eigenfunction of ${\cal P}^{(N)}$. We surmize that the spectral
determinant factors as
\begin{equation}
d(z) = \prod_{i=1}^\infty  d_i(z) \,,
\end{equation}
with one factor $d_i$ for the family $\{k\}_i$ of periodic orbits
showing up in the $i$th eigenfunction. Each such factor $d_i(z)$ is then
calculated separately with the above expressions. Obviously this variant
of the cycle expansion reproduces the phases of the resonances exactly
since these are again directly determined by lengths of orbits: if $p$
is the smallest period used in $d_i(z)$, the polynomial can as well be
written as a polynomial in $z^p$ thus allowing the zeros to have the
phases of the $p$th roots of unity. Needless to say, the notorious
proliferation of periodic orbits with growing length causes more and
more difficulties for the cycle expansion for $p$-families of resonances
with increasing $p$.

We first turn to the top with $\tau=10$. Here, unstable orbits with
periods 1,2, and 4 and the corresponding $p$-families of resonances are
easily detected. In Table \ref{tableB} frozen eigenvalues found at the
resolution $l_{{\rm max}}=60$ are listed in the first column. The
columns to the right show the corresponding results of the cycle
expansion up to order $n_{{\rm max}}=1, 2, 4$. The total number of
orbits used in each case is given in curly brackets. For instance, our
cycle
expansion of the resonance $0.81$ involves three orbits of lengths 
$1$, $2$ and $4$. The resonance $-0.7510$ as well as its phase-related
partners $-0.0079\pm {\rm i} 0.7517$ can be approximately reproduced
using four orbits of length $4$. 
For the real positive eigenvalue $0.7470$ (not listed in table 
\ref{tableB}) completing this quartet of eigenvalues the eigenfunction 
is also localized in other regions of phase space and thus  further orbits 
with periods  different from $p=4$ would have to be included in the cycle 
expansion. The first repetition of the single
period-2 orbit contributing to the resonance $0.6597$ in table 
\ref{tableB} gives an almost diverging contribution to the spectral 
determinant, thus hindering its expansion to a higher order.

The top with $\tau = 10.2$ offers itself for a study of complex frozen
eigenvalues with phases near the fourth and sixth roots of unity. Even
though these two groups of eigenvalues are almost equal in modulus, the
corresponding eigenfunctions are localized in different regions of phase
space, and therefore the two examples turn out independent (see figures 
\ref{figure4}b,c). 
The reason why we leave aside the eigenvalues on the real axis is that 
their eigenfunctions show localization on many orbits of different, 
higher periods which cannot all be identified 
(see figures \ref{figure4}a,d). 
For the resonances $-0.0058\pm {\rm i}0.7080$ the spectral determinant 
of order $n_{{\rm max}}=4$, calculated from six periodic orbits of 
length $4$, yields $\pm {\rm i}0.8150$. In the next higher order 
$n_{{\rm max}} = 8$ another $9$ orbits of primitive length $8$ must 
be taken into account with the resulting approximants $\pm {\rm i}0.7678$. 
For the resonances $0.3550\pm {\rm i} 0.6199$ and 
$-0.3388 \pm {\rm i}0.6243$ with phases near the sixth roots of unity 
$12$ orbits of primitive length $6$ contribute to the spectral determinant 
in lowest order. These orbits are shown in figure \ref{figure6}a. 
In lowest order the resulting six approximants of modulus $0.8599$ still 
differ by about $20\%$ from the respective moduli $0.7144$ and $0.7103$. 
These results are compiled in Table \ref{tableC}.

\section{Resonances versus correlation decay}

To check on the physical meaning of frozen eigenvalues with moduli
smaller than unity as Frobenius-Perron resonances we propose to compare
their moduli with rates of correlation decay. In a numerical experiment
we investigate the decay of the correlator
\begin{equation}
C(n)  = 
\frac{\langle \rho(n) \rho(0) \rangle - \langle \rho(\infty) \rho(0)
\rangle} {\langle \rho(0)^2 \rangle - \langle \rho(\infty)
\rho(0)
\rangle}\;.
\end{equation}
The initial densities $\rho(0)$ are chosen as characteristic functions
on a grid of $2000\times2000$ cells in phase space. Each cell is
represented by its central point $(q,p)_c$, and its temporal successor
after one time step is the cell containing $M((q,p)_c)$. After $n$
iterations we obtain $\rho(n)$. From the phase-space average $\langle
\rho(n) \rho(0) \rangle$ we subtract the correlations remaining at
$n=\infty$ due to the compactness of the phase space, and subsequently
normalize.
 
\begin{figure}
\epsfxsize=0.5\textwidth
\epsffile{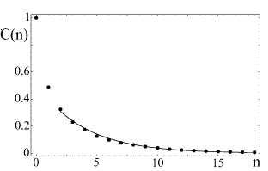}
\caption{Decay of $C(n)$ (dots) with $\rho(0)$ chosen according to the 
eigenfunction shown in figure \protect\ref{figure4}b 
(also \protect\ref{figure5}b). The numerical fit (line) yields a decay 
factor $0.7706$.}
\label{figure7}
\end{figure}

Depending on the choice of $\rho(0)$ different long-time decays are
observable. 
We choose $\rho(0)$ as covering the regions where the
hyperbolic orbits relevant for a given resonance are situated. The
long-time decay turns out insensitive to the exact choice of $\rho(0)$,
provided $\rho(0)$ does not extend to phase-space regions supporting
different eigenfunctions with frozen eigenvalues of larger moduli.
Since $\rho(0)$ is real and positive, and positivity is
preserved under the time evolution, we find positive real decay
factors $C(n+1)/C(n)$.
Indeed, after two or three time steps these decay factors are
in rather good agrement with the modulus of the corresponding
nonunimodular eigenvalue as well as with the result of our cycle expansion.
Tables \ref{tableB} and \ref{tableC} list the decay factors we obtained
from a numerical fit of $C(n)$ from $n=2$ to $n=18$, together with the
moduli of the respective resonances. Figure \ref{figure7} shows the
decay of $C(n)$ (dots), with $\rho(0)$ chosen according to the
eigenfunction shown in figure \ref{figure4}b. The numerical fit (line)
yields a decay factor $0.7706$, to be compared with $0.7103$ as the
modulus of the eigenvalue.

\section{conclusions}

In conclusion, we have presented a method to determine Frobenius-Perron
resonances and eigenvalues which is applicable to systems with a mixed
phase space. Resonances as well as eigenvalues are identified as frozen
eigenvalues of a truncated propagator matrix ${\cal P}^{(N)}$ in a
Hilbert space of phase-space functions. The corresponding eigenfunctions
of ${\cal P}^{(N)}$ are strongly localized on the associated phase-space
structures, elliptic islands for eigenvalues and unstable manifolds of
hyperbolic periodic orbits for resonances. The expected characteristics
of the resonance eigenfunctions, an infinite hierarchy of structures and
the distinct supports of foreward and backward time evolution, are
clearly visible. The resonances obtained can be reproduced by a variant
of the cycle expansion in which only the orbits that are identified in
the eigenfunctions are taken into account. Finally, the correlation
decay within a phase-space region that an eigenfunction is localized in
is well predicted by the corresponding resonance. The cycle expansion as
well as the observation of correlation decay may be used to check the
accuracy of the resonances obtained as frozen eigenvalues. An accurate
knowledge of the relevant resonances of mixed dynamics at a certain
phase-space resolution is a desirable piece of information for many
further investigations such as the role of classical resonances in
quantum dynamics.

We are grateful to Shmuel Fishman for discussions initiating as well as
accompanying this work.  Support by the Sonderforschungsbereich 
``Unordnung und gro{\ss}e Fluktuationen'' and the Minerva foundation 
is thankfully acknowledged.

\section{Appendix A: Matrix elements of the Frobenius-Perron operator}

We here derive the matrix elements (\ref{eq5}) of the Frobenius-Perron 
operator ${\cal P}$. It is convenient to use the representation of 
$Y_{lm}$ in terms of associated Legendre Polynomials $P_{l}^m(x)$ 
\cite{Biedenharn},
\begin{equation}
Y_{lm}(q,p) = (-1)^m \left[\frac{2l+1}{4\pi}\frac{(l-m)!}{(l+m)!}
\right]^{\frac{1}{2}}\, P_l^m(p) \; {\rm e}^{{\rm i} m q}\, ,
\label{ap2}
\end{equation}

For rotations $R_y, R_z$ and torsion $T_z$ on the unit sphere
the explicit forms for $M$ read
\begin{eqnarray}
R_y(\beta_y)  :
{q \choose p} & \longmapsto &
{\arg \left(\sqrt{1-p^2}\, (\cos \beta_y \cos q
+ {\rm i} \sin q ) + p \sin \beta_y  \right)
\choose p \cos \beta_y - \sqrt{1-p^2}\sin\beta_y \cos q }
\nonumber \\
R_z(\beta_z)  :
{q \choose p} & \longmapsto &
{q+\beta_z
\choose p}
\label{ap4} \\
T_z(\tau)  :
{q \choose p}& \longmapsto &
{q+\tau p
\choose p} \; . \nonumber
\end{eqnarray}

The above immediately implies the intuitive result for the $z$-rotation
\begin{equation}
Y_{l m}\left(R_z^{-1}(q,p)\right) = {\rm e}^{-{\rm i} m \beta_z}
Y_{lm}(q,p) \,.
\label{ap6}
\end{equation} 
Due to the orthonormality of the spherical harmonics we then find the
Frobenius-Perron operator represented by the diagonal  matrix
\begin{equation}
\exp({\cal L}_{R_z})_{l_1 m_1,l_2 m_2} =
\delta_{l_1 l_2}\; \delta_{m_1 m_2} \;{\rm e}^{-{\rm i} m_2 \beta_z} \,.
\label{ap7}
\end{equation}

Obviously and intuitively, $\exp({\cal L}_{R_z})$ is identical to the
quantum mechanical $(\hbar=1)$ rotation matrix $\langle l_1 m_1|
\exp(-{\rm i} \beta_z L_z) |l_2 m_2\rangle$ involving the $L_z$-component 
of the angular momentum operator $\vec{L} = (L_x, L_y, L_z)^T$ and the
eigenbasis $Y_{lm} = |l m\rangle$ with $\vec{L}^2 |lm\rangle = l(l+1)
|lm\rangle$ and $L_z |lm\rangle = m |lm\rangle$. Since this formal
identity holds for the $z$-rotation it must also hold for the
$y$-rotation. Thus for $\exp({\cal L}_{R_y})$ we must take the
representation of the quantum operator $\exp(-{\rm i} \beta_y L_y)$,
well known as Wigner's ${\rm d}$-matrix,
\begin{equation}
\exp({\cal L}_{R_y})_{l_1 m_1,l_2 m_2} = \langle l_1 m_1|\exp(-{\rm i}
\beta_y L_y) |l_2 m_2\rangle = \delta_{l_1 l_2}\; 
{\rm d}^{l_1}_{m_1 m_2}(\beta_y) \, ,
\label{ap8}
\end{equation}
which can be obtained from recursion relations \cite{Braun}. This matrix
is still diagonal in $l_1, l_2$. In fact, the Frobenius-Perron matrix for
any rotation is diagonal in that pair of indices. An immeditate
implication is that pure rotations do not couple the subspace of
$Y_{lm}$ with $l\le l_{{\rm max}}$ to its complement. For the integrable
dynamics of rotation no probability is dissipated from the subspace and
the spectrum of ${\cal P}^{(N)}$ lies exactly on the unit circle.

For the $z$-torsion  we make the replacement
\begin{equation}
Y_{lm}(T_z^{-1}(q,p))  = {\rm e}^{-{\rm i}\tau m p} \; Y_{lm}(q,p)
\label{ap9}
\end{equation} 
in the phase-space integral (\ref{eq5}) and then use the
expansion \cite{Gradshteyn}
\begin{equation}
{\rm e}^{-{\rm i} k \cos\theta} = \sum_{l=0}^\infty (-{\rm i})^l
\sqrt{4\pi (2l+1)} \; j_l(k) \; 
Y_{l 0}(\varphi, \cos\theta) \, ,
\label{ap10}
\end{equation}
which involves the spherical Bessel function $j_l(k)$. The
parity $Y^\ast_{l_1 m_1} = (-1)^{m_1} Y_{l_1,-m_1}$
helps us on to
\begin{equation}
\exp({\cal L}_{T_z})_{l_1 m_1,l_2 m_2} = (-1)^{m_1} \sum_{l=0}^\infty
(-{\rm i})^l \sqrt{4\pi (2l+1)} j_l(m_1\tau) \int_\Omega
{\rm d}q\;{\rm d}p\;  \; Y_{l 0}\; Y_{l_1,-m_1} \; Y_{l_2 m_2} \, .
\label{ap11}
\end{equation}
We here meet Gaunt's integral\cite{Gaunt} over three spherical harmonics
which can be written in terms of two Clebsch-Gordan coefficients,
\begin{equation}
\int_\Omega{\rm d}q\;{\rm d}p\;\;Y_{l m}^\ast\;Y_{l_1 m_1}\;Y_{l_2 m_2} =
{\left[ \frac{(2l_1+1)(2l_2+1)}{4\pi (2l+1)} \right]}^{\frac{1}{2}}  \; 
C_{0\, 0\, 0}^{l_1\, l_2\, l} \;C_{m_1 m_2 m}^{l_1\,\, l_2\,\, l} \, .
\label{ap12}
\end{equation}
Finally, using the following properties of Clebsch-Gordan coefficients,
\begin{eqnarray}
C_{m_1 m_2 m}^{l_1\,\, l_2\,\, l} & = & 0 \quad \mbox{if} \,\,\, l
\notin \{(l_1+l_2), (l_1+l_2)-1, \dots ,
|l_1-l_2|\} \, , \nonumber \\
C_{m_1 m_2 m}^{l_1\,\, l_2\,\, l} & = & 0 \quad \mbox{if} \,\,\,
m_1+m_2 \ne m \, , \label{ap14} \\
C_{0\, 0 \, 0}^{l_1\, l_2\, l} & = & 0 \quad \mbox{if} \,\,\,
l_1+l_2+l \,\,\, \mbox{odd} \nonumber
\end{eqnarray}  
the infinite sum in (\ref{ap11}) can be reduced to a finite one,
\begin{equation}
\exp({\cal L}_{T_z})_{l_1 m_1,l_2 m_2} =  \delta_{m_1 m_2}
(-1)^{m_1} \sqrt{(2l_1+1)(2l_2+1) } \sum_{l=|l_1-l_2|}^{l_1+l_2}
(-{\rm i})^l \; j_l(m_1\tau) \; C_{0\, 0\, 0}^{l_1\, l_2\, l} \;
C_{-m_1 m_2 0}^{l_1\,\, l_2\,\, l} \, .
\label{ap16}
\end{equation}
It is worth noting that due to the properties (\ref{ap14}) the matrix
elements (\ref{ap16}) are real if $l_1+l_2$ is even or purely imaginary
if $l_1+l_2$ is odd. Furthermore they are symmetric with respect to
$l_1$ and $l_2$.

\section{Appendix B: Recursion relation for the torsion matrix elements}

A recursion relation similar to the recursion relation used to calculate
Wigner's ${\rm d}$-matrix can be derived for the torsion matrix elements. 
For simplicity we consider the integral
\begin{equation}
t_{z}(l', l, m) = \int_{-1}^1 {\rm d}u\; P_{l'}^m(u) \,
{\rm e}^{{\rm i} \tau m u} \,  P_{l}^m(u) \,
\label{ap17}
\end{equation}
which is related to the torsion matrix elements by
\begin{equation}
\exp({\cal L}_{T_z})_{l' m l m} = \frac{1}{2} {\left[
\frac{(2l'+1) (l'-m)!}{(l'+m)!} \frac{(2l+1) (l-m)!}{(l+m)!}
\right]}^{\frac{1}{2}} \,\; t_{z}(l', l, m) \, .
\label{ap18}
\end{equation}

The associated Legendre polynomials obey the differential equation
\begin{equation}
D P_l^m(u) \equiv \left[(1-u^2) \frac{{\rm d}^2}{{\rm d} u^2}
- 2 u \frac{{\rm d}}{{\rm d} u} + l (l+1) - \frac{m^2}{1-u^2}
\right]\; P_l^m(u) = 0 \, .
\label{ap19}
\end{equation}
This suggests to draw a recursion relation from
\begin{equation}
\int_{-1}^1 {\rm d} u \, \left[ D \, P_{l'}^m(u) \right]\,
P_{l}^m(u) \, {\rm e}^{{\rm i} \tau m u} = 0 \, .
\label{ap20}
\end{equation}
In the case $m \ne 0$ integration by parts leads to
\begin{eqnarray}
\int_{-1}^1 {\rm d} u \, P_{l'}^m \, {\rm e}^{{\rm i} \tau m u} \,
\big[ (2 {\rm i} \tau m)\; \big( (1-u^2) \frac{{\rm d}}{{\rm d} u}
P_{l}^m - u\; P_{l}^m \big)\, + & & \label{ap21}\\
\big( l'(l'+1)-l (l+1)-(1-u^2)\; \tau^2 m^2 \big)\,
P_{l}^m\big] & = & 0 \, . \nonumber
\end{eqnarray}
Now with the help of the identity
\begin{equation}
(1-u^2) \frac{{\rm d}}{{\rm d} u} P_l^m=-luP_l^m+(l+m)P_{l-1}^m
\label{ap22}
\end{equation}
the derivative in (\ref{ap21}) can be eliminated and we arrive at
the five-step recursion relation
\begin{eqnarray}
t_{z}(l', l+2, m) =
 & & \frac{1}{\tau^2 m^2} \frac{(2l+1)(2l+3)}{(l-m+1) (l-m+2)}\;
\bigg[2{\rm i} \tau m \frac{(l+1) (l-m+1)}{2l+1}\; t_{z}(l',l+1,m) 
\nonumber \\
 & &-\Big(l' (l' +1)-l (l+1) - 2\tau^2 m^2 \frac{l (l+1)+m^2-1}{4l(l+1)-3} 
 \Big)\;
 t_{z}(l',l,m)\label{ap23} \\
 & & -2{\rm i} \tau m \frac{l(l+m)}{2l+1}\; t_{z}(l', l-1, m)
 -\tau^2 m^2\frac{(l+m)(l+m-1)}{(2l+1)(2l-1)}\;t_{z}(l',l-2,m)\bigg]\;.
 \nonumber
\end{eqnarray}
For constant indices $l'$ and $m\ne 0$ this links matrix elements with
five successive values of the index $l$. Just like the torsion matrix
elements for these indices the terms in the recursion relation have
alternately real and imaginary values. Even though the recursion
relation involves five matrix elements, only two integrals have to be
done initially. These are the nonvanishing integrals $t_{z}(l', |m|, m)$
and $t_{z}(l', |m|+1, m)$ for the smallest values of $l$. Together with
$t_{z}(l', |m|-1, m) = 0$ and $t_{z}(l', |m|-2, m) = 0$ they can be used
to start the recursion.

For $m=0$ the recursion formula does not apply, but the integral
\begin{equation}
t_{z}(l', l, 0) = \frac{2}{2l+1}\; \delta_{l l'}
\label{ap24}
\end{equation}
is trivial in this case.


\newpage

\begin{table}
\[
\begin{array}{rrrr} \hline\hline
l_{{\rm max}} = 30 & l_{{\rm max}} = 40 & l_{{\rm max}} = 50 & 
l_{{\rm max}} = 60 \\ \hline
0.7700 & 0.7688 & 0.7523 & 0.7696 \\ \hline
\begin{array}{r} 0.3075\\ \pm {\rm i}\; 0.5740\end{array} &
\begin{array}{r} 0.3429\\ \pm {\rm i}\; 0.6140\end{array} &
\begin{array}{r} 0.3523\\ \pm {\rm i}\; 0.6211\end{array} &
\begin{array}{r} 0.3550\\ \pm {\rm i}\; 0.6199\end{array} \\ \hline
\begin{array}{r} -0.3170\\ \pm {\rm i}\; 0.6003\end{array} &
\begin{array}{r} -0.3348\\ \pm {\rm i}\; 0.6272\end{array} &
\begin{array}{r} -0.3444\\ \pm {\rm i}\; 0.6283\end{array} &
\begin{array}{r} -0.3388\\ \pm {\rm i}\; 0.6243\end{array} \\ \hline
\begin{array}{r} -0.0042\\ \pm {\rm i}\; 0.7161\end{array} &
\begin{array}{r} -0.0002\\ \pm {\rm i}\; 0.7133\end{array} &
\begin{array}{r} -0.0100\\ \pm {\rm i}\; 0.6930\end{array} &
\begin{array}{r} -0.0058\\ \pm {\rm i}\; 0.7080\end{array} \\ \hline
-0.7025 & -0.7228 & -0.7155 & -0.7165 \\ \hline
0.6544 & 0.6230 & 0.6495 & 0.6480 \\ \hline
\dots & -0.5619 & -0.5753 & -0.5667 \\ \hline\hline
\end{array}
\]
\caption{Frozen nonunimodular eigenvalues for $\tau=10.2$ at 
$l_{{\rm max}} =30, 40, 50$ and $60$ that can be identified from the 
histogram shown in figure \protect\ref{figure2}b.}
\label{tableA}
\end{table}

\begin{table}
\[
\begin{array}{rrrrr} \hline \hline
l_{{\rm max}} = 60 & n_{{\rm max}} = 1  & n_{{\rm max}} = 2 & 
n_{{\rm max}} = 4 & \mbox{Decay of}\;C(n)\\ \hline
0.8103 & 0.2185 \{1\} & 0.7070 \{2\} & 0.7664 \{3\} & 0.8005\\ \hline
-0.7510 & \dots & \dots & -0.7483 \{4\} & 0.7697\\ \hline
\begin{array}{r} -0.0079\\ \pm {\rm i}\; 0.7517\end{array} & \dots & 
\dots & -0.7483 \{4\} & 0.7697 \\ \hline
0.6597 & \dots & 0.4969 \{1\} & \dots & 0.6783 \\ \hline \hline
\end{array}
\]
\caption{Left column: resonances obtained from the truncated propagator
for $\tau = 10.0$ and $l_{{\rm max}} = 60$. Middle columns: corresponding 
results from cycle expansion up to order $n_{{\rm max}} = 1,2, 4$. The 
total number of primitive orbits employed is
given in curly brackets. The eigenvalues in the second and third row  
belong to the same quartet of phase-related partners. Right column: 
Associated
decay factors by which
$C(n)$ decreases over one time step, obtained from numerical fit.}
\label{tableB}
\end{table}

\begin{table}
\[
\begin{array}{rrrrr} \hline \hline
l_{{\rm max}} = 60 & n_{{\rm max}} = 4  & n_{{\rm max}} = 6 & 
n_{{\rm max}} = 8 & \mbox{Decay of}\;C(n)  \\ \hline
\begin{array}{r} -0.0058\\ \pm {\rm i}\; 0.7080\end{array} &  
\pm {\rm i}\; 0.8150 \{6\} & \dots &
 \pm {\rm i}\; 0.7678 \{15\} & 0.7694\\ \hline  
\begin{array}{r} 0.3550\\ \pm {\rm i}\; 0.6199\end{array} & \dots & 
\begin{array}{r} 0.4299\\ \pm {\rm i}\; 0.7447\end{array} \{12\} & 
\dots & 0.7706 \\ \hline 
\begin{array}{r} -0.3388\\ \pm {\rm i}\; 0.6243\end{array} & \dots & 
\begin{array}{r} -0.4299\\ \pm {\rm i}\; 0.7447\end{array} \{12\} & 
\dots & 0.7706\\ \hline 
\end{array}
\]
\caption{Left column: resonances obtained from the truncated propagator
for $\tau = 10.2$ and $l_{{\rm max}} = 60$. Middle columns: corresponding 
results from cycle expansion up to order $n_{{\rm max}} = 4,6,8$. The 
total number of primitive orbits employed is given in curly brackets. 
The eigenvalues in the second and third row belong to the same sextet of
phase-related partners. Right column: Associated decay factors by which
$C(n)$ decreases over one time step, obtained from a numerical fit.}
\label{tableC}
\end{table}

\newpage

\newpage

\newpage

\newpage

\newpage

\newpage

\newpage

\begin{thebibliography}{99}
\bibitem{Ruelle}
D. Ruelle, Phys. Rev. Lett. {\bf 56}, 405 (1986);
D. Ruelle, J. Stat. Phys. {\bf 44}, 281 (1986).
\bibitem{Pollicott}
M. Pollicott, Invent. Math. {\bf 81}, 415 (1985)
\bibitem{Us}
F. Haake, C. Manderfeld, J. Weber, and P. A. Braun, in preparation
\bibitem{Pance} 
K.Pance, W. Lu, S. Sridhar, Phys. Rev. Lett. {\bf 85}, 2737 (2000)
\bibitem{Altshuler} A.~V. Andreev and B.~L. Altshuler, Phys. Rev. Lett. 
{\bf 75}, 902 (1995);
O. Agam, B.~L. Altshuler, and A.~V. Andreev, Phys. Rev. Lett. {\bf 75}, 
4389 (1995);
A.~V. Andreev,O. Agam, B.~D. Simons, and B.~L. Altshuler, Phys. Rev. Lett. 
{\bf 76}, 3947 (1996);
A.~V. Andreev, B.~D. Simons, O. Agam, and B.~L. Altshuler,  Nuclear 
Physics B, {\bf 482}, 536 (1996).
\bibitem{Zirnbauer}
M.~R. Zirnbauer in: I.V. Lerner,
J.P. Keating, and D.E. Khmelnitskii (eds.), {\em Supersymmetry and Trace
Formulae:  Chaos and Disorder}
(Kluwer Academic, New York, 1999)
\bibitem{Reed}
M. Reed and B. Simon, {\em Methods of Modern Mathematical Physics IV:
Analysis of Operators} (Academic Press, New York, 1978).
\bibitem{Fishman} S. Fishman in: I.V. Lerner, J.P. Keating, and D.E.
Khmelnitskii (eds.),
{\em Supersymmetry and Trace Formulae:  Chaos and Disorder} (Kluwer
Academic, New York, 1999)
\bibitem{Khodas} M. Khodas and S. Fishman, Phys. Rev. Lett. {\bf 84}, 
2837 (2000);
M. Khodas and S. Fishman, Phys. Rev. E {\bf 62}, 4769 (2000).
\bibitem{Hasegawa}
H.~H. Hasegawa und W.~C. Saphir, Phys. Rev. A {\bf 46}, 7401 (1992)
\bibitem{Baladi}
V. Baladi, J.-P. Eckmann, and D. Ruelle, Nonlinearity {\bf 2}, 119 (1989)
\bibitem{Weber}
J. Weber, F. Haake, and P. \v{S}eba, Phys. Rev. Lett. {\bf 85}, 3620 (2000)
\bibitem{Haake}
F. Haake, {\em Quantum Signatures of Chaos, 2nd ed.}
(Springer, Berlin, 2001)
\bibitem{Titchmarsh}
E.~C. Titchmarsh, Proc. Roy. Soc. London Ser. A {\bf 210}, 30 (1951)
\bibitem{Kato} 
T. Kato, {\em Perturbation Theory for Linear Operators} 
(Springer, Berlin, 1995)
\bibitem{Cvitanovic}
P. Cvitanovi\'{c} et al., {\em Classical and Quantum Chaos: A Cyclist
Treatise},
(Nils Bohr Institute, Copenhagen, 1999) (www.nbi.dk/ChaosBook/)
\bibitem{Christiansen}
F. Christiansen, G. Paladin, and H.~H. Rugh, Phys. Rev. Lett. {\bf 65}, 
2087 (1990)
\bibitem{Braun}
P.~A. Braun, P. Gerwinski, F. Haake, and H. Schomerus, Z. Phys. B 
{\bf 100}, 115 (1996)
\bibitem{Biedenharn}
L.~C. Biedenharn and J.~D. Louck, {\em Encyclopedia of Mathematics and 
Its Applications: Angular 
Momentum in Quantum Physics}
(Addison-Wesley, Reading, 1981)
\bibitem{Gradshteyn}
I.~S. Gradshteyn and I.~M. Ryzhik, {\em Table of Series, Integrals, and 
Products}
(Academic Press, San Diego, 1994)
\bibitem{Gaunt}
J. A. Gaunt, Trans. Roy. Soc. A {\bf 228}, 151 (1929)

\end{thebibliography}
\end{document}